# Quantifying residual stress in Helium implanted surfaces and its implication for blistering.

P. Hosemann[ab*], M. Sebastiani[c], M. Z. Mughal[c], X. Huang[a], A. Scott[a], M. Balooch[a]

[A] University of California at Berkeley, Department of Nuclear Engineering, CA, Berkeley, 94720, USA
[B] Lawrence Berkeley National Laboratory, Material Science Division, CA, Berkeley, 94720
[C] Roma Tre University, Engineering Department, Via della Vasca Navale 79, 00146 Rome, Italy

*Corresponding Author

**Abstract:**

Helium implantation in surfaces is of interest for plasma facing materials and other nuclear applications. Vanadium as both a representative bcc material and a material relevant for fusion applications is implanted using a Helium ion beam microscope, and the resulting swelling and nanomechanical properties are quantified. These values are put in correlation to data obtained from micro residual stress measurements using a focused ion beam based ring-core technique. We found that the swelling measured is similar to literature values. Further we are able to measure the surface stress caused by the implantation and find it approaches the yield strength of the material at blistering doses.

The simple calculations performed in the present work, along with several geometrical considerations deduced from experimental results confirm the driving force for blister formation comes from bulging resulting mainly from gas pressure buildup, rather than solely stress induced buckling.

**Graphical Abstract:**

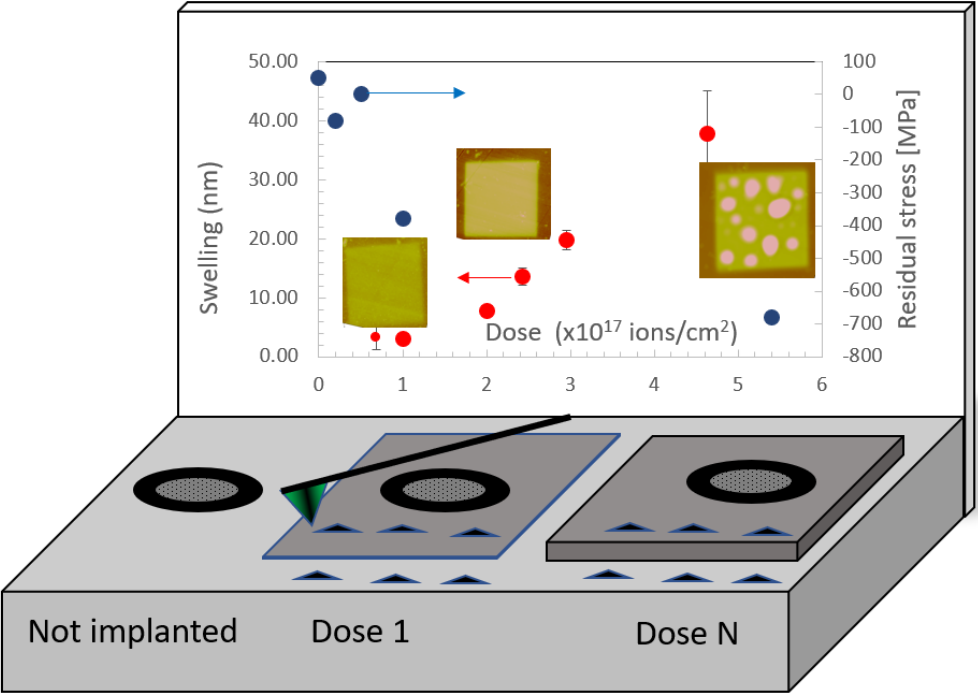

**Introduction:**

The effect of Helium in materials is an important factor in nuclear fission and fusion applications. Helium build up occurs due to (n, α) reactions in materials or via direct impact, such as surface plasma interactions in fusion devices. The produced Helium is highly insoluble in metals and forms gas bubbles, which are the subject of numerous studies on actual in-service materials such as Inc X-750 [1, 2]. Within a certain range of temperatures and Helium production rates, smaller bubbles often develop in a periodic arrangement where Helium bubbles self-order following the host lattice structure and orientation [3]. These Three Dimensional (3D) ordered bubble structures are called superlattices and have been studied for decades [4,5]. Recent work showed that these lattices can be maintained during twinning processes but not during ordinary plastic deformation [6]. The trapped Helium leads to a volume change in the material and therefore to gas swelling. The gas swelling is a function of bubble size and distribution. Initially the bubbles grow by the addition of vacancies which are also present during irradiation while subsequent bubble growth occurs due to loop punching [7, 8]. If Helium content increases, it has been found that blisters can develop at the surface [9]. Recently it has been shown that the original proposed blistering mechanism of crack formation due to linking up of bubbles occurs [9]. While these mechanisms are relatively well understood the question remains to what extent stresses are contributing to the swelling of the material and force the material to a plastic deformation (growth) and eventual crack propagation in the blistering regime. This work intends to shine light into the micro-scale residual stresses that can develop in surface near regions due to Helium implantation. Vanadium alloys are considered for self-cooled Li concepts [10]. Vanadium alloys, especially the V-Cr-Ti system have favorable mechanical and physical properties enabling the development of future fusion power systems [11, 12]. Furthermore, Vanadium lends itself well for basic science studies as a low-density representative of bcc alloys, enabling this study and the utilization of the high precision Helium ion beam microscope (HIM) as an implantation technique [13]. We consider Vanadium as a model material to study the physics of blistering, similar to Kaminsky and Das [14-16]. Other bcc systems such as Niobium have been utilized in the past where clear correlations of dose, blister diameter, and blister layer thickness were found. These studies utilized macroscopic implantations at beam energies of 250keV and higher [17,18]. While the beam energy is different than the study performed here the flux and in a HIM are also significantly different. The present work expands upon previous results by using localized low energy Helium implantation with precise control of dose, and brings the resulting mechanical property changes into discussion.

**Results and Discussion:**

The atomic force microscopy measurements conducted on the implanted regions (figure 1) allowed for detailed swelling measurements. It was found that the sample showed increased swelling with increased dose, up to 37.8 nm at the highest dose irradiated ($4.63 \times 10^{17}$ ions/cm$^2$). As can be seen in Figure 2, this is also the dose at which the formation of blisters starts to occur. The material exhibited a lower blistering threshold than values previously reported for Copper [19] where blistering occurred at around $1 \times 10^{18}$ ions/cm$^2$. However, the amount of swelling is similar to what was reported for Copper. While the previous works from Y. Yang [19] on Copper focused on quantifying the swelling via TEM and AFM through local implantations utilizing the HIM technique, the present work focuses on the driving forces behind it up to the point where blistering occurs.

The insertion of bubbles into the bulk material should lead to a uniform expansion in all three dimensions resulting in homogeneous swelling. However, because the region is surrounded by non-implanted material the expansion is constrained to only allow swelling at the free surface. Therefore, the height increase is expected to be larger than that of homogenous expansion perpendicular to the surface. Simple Stopping Range of Ion in Matter (SRIM) [20] calculations using 40eV displacement energy in the Kinchin-Pease (K-P) mode shows a maximum penetration depth of ~200 nm (figure 1). The calculation does not take into account the channeling phenomena that might occur and therefore represent the lower end of the penetration depth. Assuming homogeneous Helium distribution throughout the depth of implant, the swelling in percent is approximated as:

$$S[\%] = \frac{\Delta V}{V} 100 = \frac{\Delta h}{h} 100 \qquad \text{(Equation 1)}$$

where h is the total beam penetration depth calculated by SRIM [20] (maximum depth of Helium ions in the material) with the change of height, $\Delta h$, defined as swelling, measured by AFM using the average height (excluding blisters) measured via the line scan feature. Experimental results suggest 18% swelling at the maximum dose of $4.6 \times 10^{17}$ ions/cm$^2$. A complete plot of swelling as a function of dose can be found in Figure 3a. The material can only flow towards the free surface due to the remaining bulk material constraints and the tremendous stresses that build up due to the density change. Since the ductile to brittle transition temperature of Vanadium is below room temperature (irradiation temperature) [12] we can assume that multiple slip systems are active and plastic deformation is not hindered.

Figure 3b shows the relaxation strain profiles versus dose along with the average residual stress derived from the focused ion beam digital image correlation (FIB-DIC) milling procedure. The Vanadium single crystal sample (control) shows insignificant tensile residual stress within the experimental error. A recent study has shown that mechanical polishing does not induce any significant residual stress and hence the sample can be considered as stress free [21].

Up to $5\times10^{16}$ ions/cm$^2$ no measurable residual stress is found, within the experimental error, even though the AFM suggests a swelling of 6 nm. However, at $1\times10^{17}$ ions/cm$^2$ and $4.6\times10^{17}$ ions/cm$^2$, leading to 8.5 nm and 37.8 nm swelling respectively, significant compressive stresses of ~299 MPa and ~507 MPa are deduced from the strain measurement. These numbers can now be compared to the yield strength of Vanadium. The values measured here for residual stress are higher than the yield strength of pure Vanadium (150-300MPa) [22, 23] reported in the literature. Since the FIB-DIC method can only measure elastic residual stress (which, indeed, are always elastic by nature), we must assume that the yield strength of the implanted material is increased due to the creation of defects, as otherwise we would not observe such high elastic residual stresses. We also note that elastic strains and elastic constants on Helium implanted bcc metals such as Tungsten [24] were measured using X-ray diffraction and Surface Acoustic Wave (SAW). Nanoindentation tests conducted in this study confirm that Helium implantation increases the hardness and reduces the elastic modulus, as is clearly shown in Figure 4(a) and 4(b). Measured at a depth of 50 nm, the hardness increased from around 2GPa to around 5GPa after an irradiation of $2\times10^{17}$ ions/cm$^2$, so we may expect a proportional increase in yield strength. We chose 50 nm as the depth of comparison because the plastic deformation zone beneath the indenter likely extends at least 4 times its depth, and therefore samples the entire implanted region. The hardness increase is consistent with those reported in [13] where also a 2.5GPa hardness increase was found with 1.1at% He content. Unfortunately, it is difficult to make comparisons beyond a dose of $2\times10^{17}$ ions/cm$^2$. The nanoindentation hardness values are reduced beyond this dose as the material is compromised by significant decreases in density and severe blistering, rendering a foam-like structure within the implanted region.

It is well-established that micro scale deformations suffer from size effects that complicate their extrapolation to meso and macro scale deformations [25,26]. S. Min Han et al [27] performed a series of nanopillar measurements as a function of pillar size on Vanadium specifically and found that 200 nm sized pillars (the same size as our ion beam penetration) experienced a 2% flow stress of around 1000MPa, which is higher than the residual stresses measured here. Of course, one would need to consider the plastic deformation zone around an indenter and compare ~40 nm deep indents (~2GPa unirradiated) with this data. Applying Tabor's rule gives an estimated yield strength of 680 MPa, which is lower than the uniaxial nano pillar compression test from [27], but this comparison may not be valid. Furthermore, previous work suggests that on irradiated materials the scaling laws don't apply since the resistance to plastic deformation is driven by obstacle interactions, rather than dislocation source interactions as in unirradiated materials [28].

Recent work by F. Allen [9] shows that blistering and cracking occurs due to the linking up of bubbles to nucleate nano-cracks that ultimately develop into micro-cracks. We may assume the same mechanism applies for Vanadium here. Measuring residual strain via the ring-core drilling method can help explain the stresses that drive the nucleation and growth of cracks, which in combination with gas pressure build up drives blistering.

Of course, the question can be raised whether a relatively ductile material (at room temperature) like Vanadium can be compared to a material that is brittle under similar conditions. We do argue that even in ductile materials it is the joining of small bubbles together that leads to cavity growth and ultimately blistering driven by the internal gas pressure and residual stress. The biggest difference observed is that we do not see "popping out" of blisters from the surface like in Tungsten, where the shell of the blister can be visibly cracked, and in some cases entirely removed.

The significantly reduced occurrence of this behavior in Vanadium may result from stabilized crack growth and plastic deformation that inhibits crack propagation through the surface.

2D stress surface buckling is common in various areas of research such as thin films [29]. In most instances the underlying material excites a compressive stress on the thin film. Here however, we have the incorporation of gas in the thin layer and therefore different aspects may need to be considered. Following the formulation by Eren et. al. [30] on Hydrogen blister formation in Hydrogen-implanted Gold film:

The post-lateral stress in the surface is described as

$$\sigma = 4.88 k E_r \left(\frac{h}{c}\right)^2 + \sigma_c \qquad \text{Equation 2}$$

where $\sigma_c$ is the classical critical stress for buckling:

$$\sigma_c = 4.88 E_r \left(\frac{t}{c}\right)^2 \qquad \text{Equation 3}$$

and

$$k = 0.2473[(1 + \nu_{13}) + 0.2331(1 - \nu_{13}^2)] \qquad \text{Equation 4}$$

$$E_r = \frac{E_n}{(1 - \nu_{13}^2)} \qquad \text{Equation 5}$$

where $E_r$ is the reduced elastic modulus, $E_n$ is the elastic modulus in the normal orientation to the surface, h is the height of the blister, c is the diameter of the blister, t is the thickness of the blister shell (the layer on top of the cavity), and $\nu_{13}$ is the Poisson ratio. We find that the stress here based on the c/h ratio in Figure 5a using equation 5 and 6 is 11.2 GPa. This isorders of magnitude higher than residual stress (507 MPa from Fig. 3) and therefore not realistic and the residual stress cannot be the sole reason for blistering. However, this approach neglects the gas pressure originating from the implanted Helium. The model based on bulging due to internal pressure, P, by M.K. Small [31] which considers the condition of residual stress plus the gas pressure in the cavities is more appropriate:

$$P = \frac{8Yth^3}{3(\frac{c}{2})^4} + \frac{4\sigma th}{(\frac{c}{2})^2} \qquad \text{Equation 6}$$

where σ is the residual stress measured here, t is the thickness of the shell, and Y being the biaxial modulus *E/(1-v)*. This model considers that both the gas pressure in the cavities and the residual stress both contribute to the blister formation. From the geometry deduced by AFM scans of blisters and an elastic modulus of ~100 GPa as measured by indentation, the first term in equation 6 is estimated as ~ 427 MPa. The second term, as the residual stress calculated from the present work, adds ~81 MPa to the equilibrium pressure estimation to total of 508 MPa. We use 100GPa since the elastic modulus of the material is reduced as a function of Helium content. This is also shown in Figure 4(b) where a dose of 3-5 x $10^{17}$ He ions/cm² lead to an elastic modulus of about 100GPa. This reduction has previously been observed in other materials [6] and is explained by the gas filled volume caused by bubble formation.

Assuming an internal pressure of 508MPa and ideal gas law, the Helium density, n, is obtained for a spherical cap geometry with radius *a*

$$n = \frac{p}{RT6}\pi h(3a^2 + h^2) \qquad \text{Equation 7}$$

This reveals a total content of around $3.7 \times 10^{10}$ Helium ions trapped within a 2μm diameter blister, assuming equilibrium conditions. We assume an open space of 300 nm (blister height h in figure 5c) relative to its surrounding) and 1.4μm open space diameter (blister diameter c subtracted by the wall thickness in figure 5c)). We can compare this to the $1.4 \times 10^9$ total amount of Helium atoms implanted in the same area (2 μm diameter) using the $4.5 \times 10^{17}$ ions/cm² sample. Further we can state that most of the Helium remains in small bubbles and does not escape in the crack. While this calculation suggests that the blisters here are below equilibrium pressure it is remarkable that these simple calculations with crude geometrical assumptions lead to only slightly different numbers.

The simple calculation above highly suggests that Helium gas is the main driving force for the final state of the blister as measured mainly through a gas pressure-induced bulging mechanism, rather than through stress-induced buckling mechanism.

A different argument for the driving force of the gas bubble pressure dominating the blistering and bulging is the fact that a small equilibrium bubble can accommodate significantly more Helium/volume than a large equilibrium bubble which is illustrated with the following gedankenexperiment. The ideal equation describing the amount (m) of gas that can be accommodated in a bubble is:

$$m = \frac{4}{3}\pi R^2 \frac{2\gamma}{kT}$$

Equation 8

With $\gamma$ being the surface tension, k Bolzman constant, T temperature and R bubble radius ( assumption ideal gas). Assuming tow bubbles were to join and create the double one can calculate the equivalent radius if two bubbles form a new sphere as $R_{equivalence} = \sqrt[3]{2} R_{bubble}$ or if n bubbles join together $R_{equivalence} = \sqrt[3]{n}\, R_{bubble}$. The illustrates that the radius change does not scale linearly with the combined volume. Considering this relationship in the equation 8 one can clearly see that a larger bubble originating from the combination of many smaller bubbles cannot accommodate the same amount of gas and therefore exceeds the equilibrium pressure leading to a rapid expansion or as in this case blistering.

**Summary:**

The study conducted here on Helium implanted Vanadium utilizes the HIM as a tool to conduct fast and efficient surface near implantations coupled with rapid characterizations such as AFM and nanoindentation to provide swelling and mechanical property data. Further, the FIB ring-core drilling method was deployed to quantify the residual stress caused by the Helium implantation. The Helium implantation causes a residual stress of about 500 MPa at $5\times10^{17}$ ions/cm$^2$ implanted dose where blistering was observed. This residual stress is close to the yield stress or fracture stress of bulk Vanadium and therefore can initiate nano-cracks, which in combination with Helium gas leakage from nearby bubbles causes the formation of micro-cracks and eventual blistering, with Helium pressure near equilibrium within the blister. In short, the simple calculations performed in the present work along with several geometrical considerations deduced from experimental results confirm the driving force for blister formation comes from bulging due mainly to the gas pressure buildup mechanism rather than solely stress induced buckling. It is interesting to note that Das and Kaminsky [15] also suggested this as being the dominant mechanism as we prove here now with actual input parameters to the residual stress components.

**Experimental:**

High purity (99.99%) single crystal Vanadium was purchased from Princeton Scientific. The crystal orientation was along the <100> axis and highly polished. The sample was irradiated utilizing the Orion Nanofab Helium Ion Beam microscope (HIM) at the Berkeley QB3 facility. The ion beam irradiation conditions were similar to those reported in [9,19,32,33] for Tungsten, Copper, Silicon Carbide, and steel, respectively. 10μm x 10μm square fields using a beam current of 93-95pA without an aperture were implanted with several different doses starting at $2\times10^{16}$ ions/cm$^2$ to $5.4\times10^{17}$ ions/cm$^2$ (15min/field at this dose). The acceleration energy was 25keV. Numerous implantations at similar doses were performed to create several implantation fields for different measurements. Some implanted fields were used for residual stress measurement, with their duplicates used for indentation and AFM. The dose levels vary slightly between the different irradiated regions. After irradiation the samples were investigated using Atomic Force Microscopy (DI IIIA) to quantify the height change due to Helium implantation as a function of dose as described in [32]. Data analysis was performed using Gwyddion software. Further, nanoindentation on all fields was conducted using the Continuous Stiffness Measurement (CSM) method with a Hysitron TI-950 nanoindenter. The tip was calibrated against fused silica.

Residual stress measurements were performed using the FIB-DIC micro-ring-core method, which has become a widespread technique for evaluating residual stress at micro-to nano-scale and has wide ranging applications in various disciplines of materials science [34,35]. The ring-core procedure was performed on an FEI Helios Nanolab 600 dual beam FIB/SEM using a current of 48 pA at 30keV. In the present work, the FIB experiments were designed in order to highlight the contribution to the residual stress from the Helium irradiated volume. Using the geometrical optimization proposed in a recent paper [36], step by step milling was carried out within the field of implantation using a semi-automated procedure. The diameter of the central stub was set equal to 4 µm and the maximum milling depth of 350 nm. Utilizing this procedure, it is possible to achieve the maximum sensitivity to surface stresses. High resolution SEM micrographs were acquired before and after each milling step using an integral of 128 images and at 50 ns dwell time while maintaining the same contrast as the reference image. A Digital Image Correlation (DIC) software was used to obtain the relaxation strain profiles from the micrographs. An equiaxial distribution of residual stresses was assumed over the region of interest and average residual stresses were calculated from the interpolation of the cumulative experimental relaxation strain at the maximum milling depth, and the use of the FEM calculated influence functions presented in previous papers [37-40]. For all calculations, a value of 100 GPa was used for modulus of elasticity

(E), from nanoindentation experiments, while a value of 0.37 was used for the Poisson's ratio (ν). Figure 2 highlights the step by step milling procedure of the sample $5\times10^{17}$ ions/cm$^2$ and $2\times10^{16}$ ions/cm$^2$ along with an overview image of the ring-core milling at various Helium implantations. For those cases where bubbles are present, the central stub was positioned in order to be between them, so to investigate the actual stress state given by the formation of the micro-voids (see figure 2a)).

**Acknowledgement:**

The authors would like to thank Dr. Frances Allen for assistance with ORION HIM operation. Support was provided by NSF-DMR Program # 1807822. All FIB activities were carried out at the Interdepartmental laboratory of electron microscopy (LIME) of University of "Roma Tre", Rome, Italy, with partial funding from the European Commission through the project OYSTER, grant number 760827, https://cordis.europa.eu/project/id/760827 , www.oyster-project.eu

**Figures:**

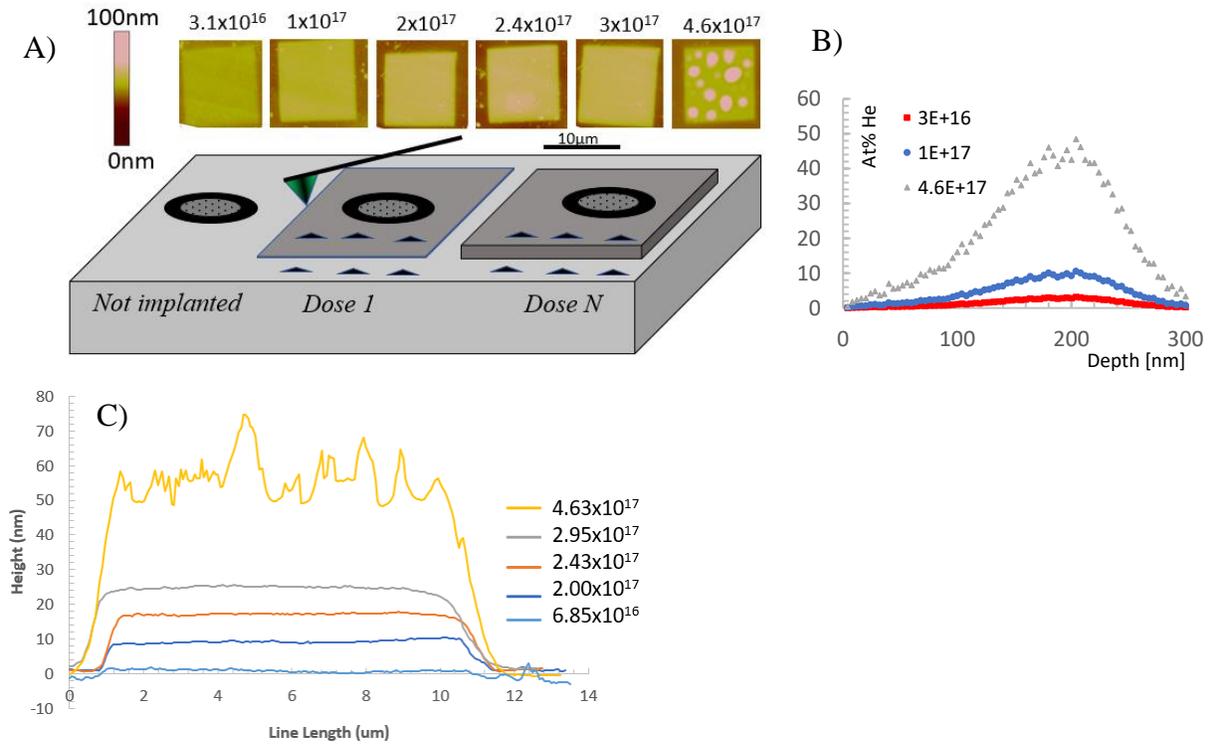

*Figure 1: Schematic representation of Helium ion implantation and associated AFM images (10x 10μm implanted areas at different doses) A) SRIM calculations of Helium ion beam implantation in Vanadium B) and line profiles from the AFM images shown B).*

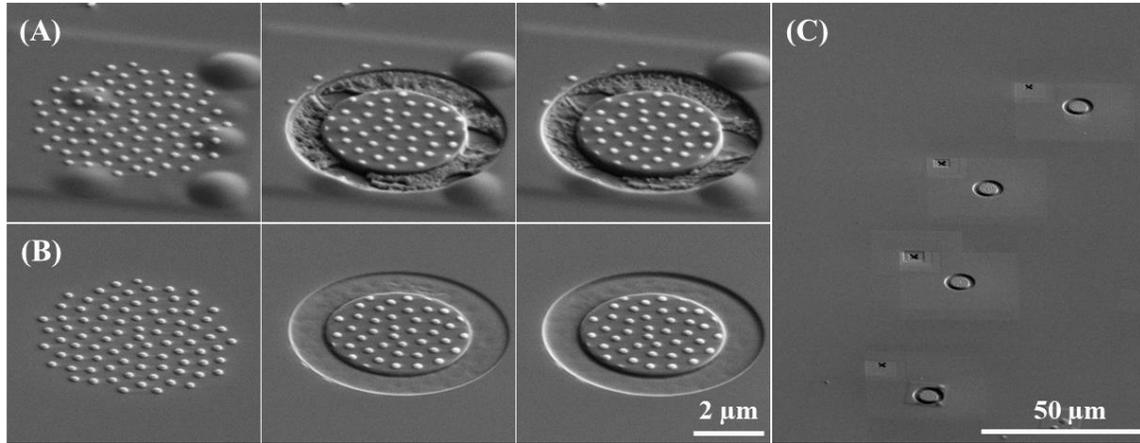

*Figure 2: (A) and (B) highlights the step by step milling procedure for sample 5x10$^{17}$ (A) and 2x10$^{16}$ (B) respectively. Surface features are platinum dots to facilitate the DIC procedure. (C) is an overview of test locations with different Helium implantation.*

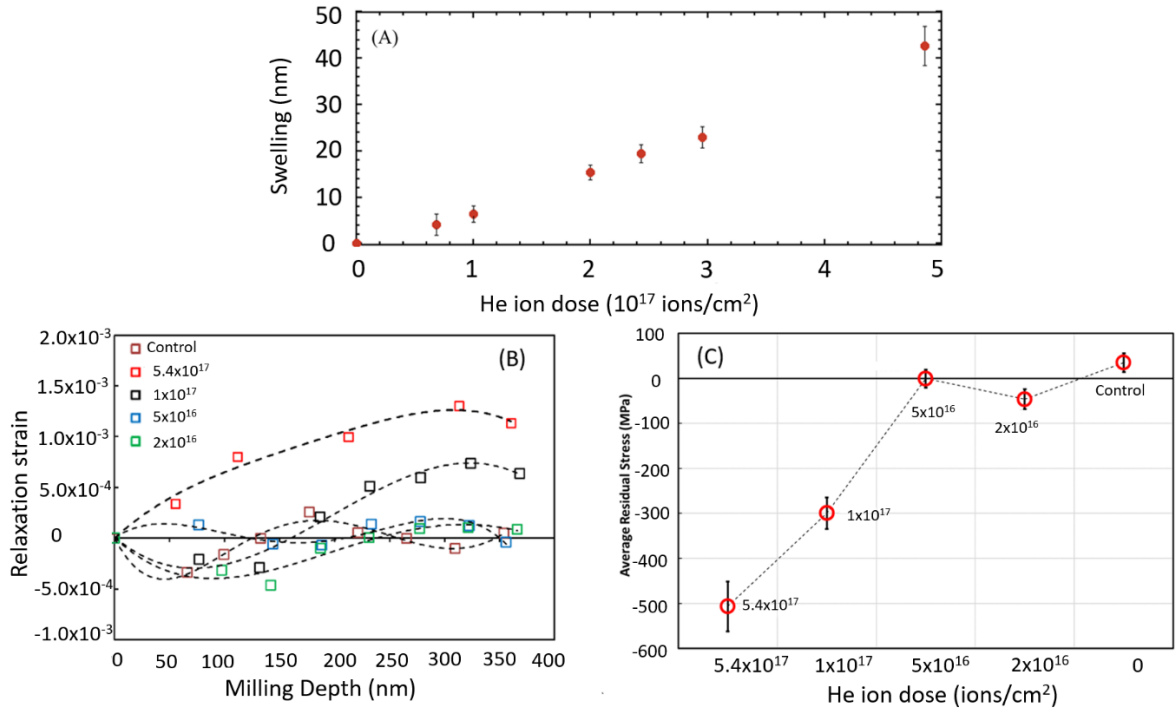

*Figure 3: Swelling as a function of dose (A) Relaxation strain as a function of milling depth (B) and average residual stress for different Helium implantation in the Vanadium single crystal C)*

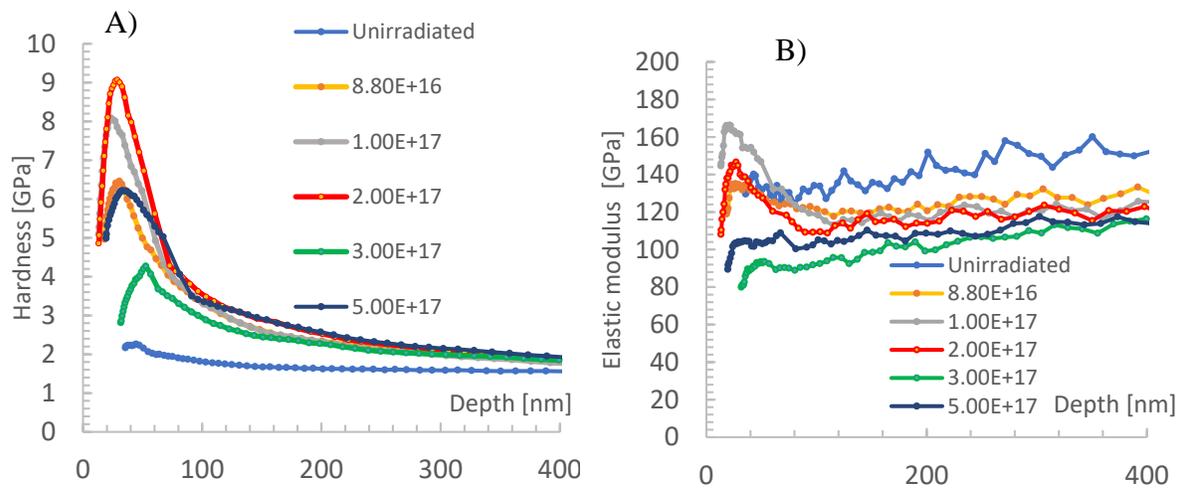

*Figure 4: Nanohardness A) and elastic modulus data B) on the Helium implanted fields.*

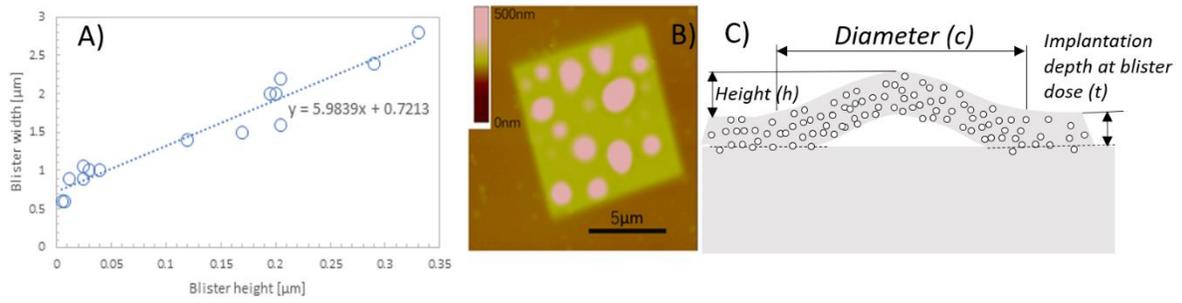

*Figure 5: Blister width and blister height ratio on the 4.6x10$^{17}$ implanted sample A) from the image to right B) and a schematic cartoon of the parameters measured from the AFM scan.*